\documentclass[english]{article}
\usepackage[T1]{fontenc}
\usepackage[latin9]{inputenc}
\usepackage{geometry}
\usepackage{amsmath}
\usepackage{amsthm}
\usepackage{amssymb}

\makeatletter
\newcommand{\lyxaddress}[1]{
	\par {\raggedright #1
	\vspace{1.4em}
	\noindent\par}
}

\@ifundefined{date}{}{\date{}}
\usepackage[all]{xy}

\usepackage{placeins}
\usepackage{lineno}
\usepackage{flushend}
\usepackage{graphicx}
\usepackage{changepage}
\usepackage{cite}
\usepackage[titletoc,title]{appendix}
\usepackage{titlesec}

\titleformat{\subsection}[runin]
  {\normalfont\large\bfseries}{\thesubsection}{1em}{}
\titleformat{\subsubsection}[runin]
  {\normalfont\bfseries}{\thesubsubsection}{1em}{}

\usepackage{babel}

\makeatother

\usepackage{babel}
\begin{document}
\title{Measures of Contextuality and Noncontextuality}
\author{Janne V. Kujala\textsuperscript{1} and Ehtibar N. Dzhafarov\textsuperscript{2}}
\maketitle

\lyxaddress{\begin{center}
\textsuperscript{1}University of Turku, janne.kujala@utu.fi\\
 \textsuperscript{2}Purdue University, ehtibar@purdue.edu 
\par\end{center}}
\begin{abstract}
We discuss three measures of the degree of contextuality in contextual
systems of dichotomous random variables. These measures are developed
within the framework of the Contextuality-by-Default (CbD) theory,
and apply to inconsistently connected systems (those with ``disturbance''
allowed). For one of these measures of contextuality, presented here
for the first time, we construct a corresponding measure of the degree
of noncontextuality in noncontextual systems. The other two CbD-based
measures do not suggest ways in which degree of noncontextuality of
a noncontextual system can be quantified. We find the same to be true
for the contextual fraction measure developed by Abramsky, Barbosa,
and Mansfield. This measure of contextuality is confined to consistently
connected systems, but CbD allows one to generalize it to arbitrary
systems.

\end{abstract}

\section{Introduction}

\subsection{}

We will consider certain \emph{measures of contextuality} (degree
of contextuality in a contextual system) and see if they can be naturally
extended into \emph{measures of noncontextuality} (degree of noncontextuality
in a noncontextual system). What we mean by an extension being ``natural''
is that it uses essentially the same mathematical construction as
the measure of contextuality being extended. Let us illustrate this
by an example. Let $\mathcal{R}$ be a system of random variables,
and let $F\left(\mathcal{R}\right)$ be a real-valued continuous functional,
in the sense that small changes in the distributions of $\mathcal{R}$
result in small changes of $F\left(\mathcal{R}\right)$. Let the following
Bell-type inequality be accepted as a definition, or derived as a
theorem: the system $\mathcal{R}$ is noncontextual if and only if
$F\left(\mathcal{R}\right)\leq0$. Suppose that in the universe of
possible systems $\mathcal{R}$ the value of $F\left(\mathcal{R}\right)$
varies on the interval $\left(a,b\right)$, with $a<0<b$. It is natural
then to consider a positive value of $F\left(\mathcal{R}\right)$
as the degree of contextuality of $\mathcal{R}$, increasing as $F\left(\mathcal{R}\right)$
increases from 0 to $b$: 
\begin{equation}
F\left(\mathcal{R}\right)>0\Longrightarrow\textnormal{CNT}=F\left(\mathcal{R}\right).
\end{equation}
Equally naturally, this measure can be extended to a measure of noncontextuality,
increasing as $F\left(\mathcal{R}\right)$ decreases from 0 to $a$:
\begin{equation}
F\left(\mathcal{R}\right)\leq0\Longrightarrow\textnormal{NCNT}=-F\left(\mathcal{R}\right).
\end{equation}
By contrast, if the functional $F\left(\mathcal{R}\right)$ varied
on an interval $\left[0,b\right)$, the degree of contextuality would
be defined as before, but it would not naturally extend to a measure
of noncontextuality: all noncontextual system would be mapped into
zero, so any extension would require ideas and principles other than
those used in the construction of the functional $F$.

\subsection{}

We will consider three contextuality measures, all based on the Contextuality-by-Default
(CbD) theory and applicable to arbitrary systems of dichotomous random
variables.\footnote{In the contemporary version of CbD \cite{DzhCerKuj2017,DzhKuj2017.2.0},
any system of random variables is to be presented in a \emph{canonical
form}, one in which each original random variable is replaced with
a set of jointly distributed dichotomous variables.} Two of our measures, $\textnormal{CNT}_{1}$ and $\textnormal{CNT}_{2}$,
are, in a well-defined sense, mirror images of each other, but we
will see that only one of them, $\textnormal{CNT}_{2}$, is naturally
extendable to a measure of noncontextuality. $\textnormal{CNT}_{2}$
for a contextual system of random variables is defined as the $L_{1}$-distance
between the surface of a certain polytope and an external point representing
the system. The points lying on or inside the polytope represent noncontextual
systems, and it is natural to define the extension of $\textnormal{CNT}_{2}$
into a measure of noncontextuality, $\textnormal{NCNT}_{2}$, as the
$L_{1}$-distance from an internal point of the polytope to its surface.
$\textnormal{CNT}_{1}$, too, can be defined as the $L_{1}$-distance
between a certain polytope and an external point representing a contextual
system. However, all noncontextual systems in this case are represented
by points lying on the surface of the polytope, as points of zero
contextuality. As a result, any extension of $\textnormal{CNT}_{1}$
into a measure of noncontextuality would require that one go beyond
the construction underlying $\textnormal{CNT}_{1}$.

\subsection{}

The third CbD-based measure, $\textnormal{CNT}_{3}$, is of a different
kind. Here, one maps the system into a certain distribution of quasiprobabilities,
numbers that sum to unity but are allowed to be negative. $\textnormal{CNT}_{3}$
is measured by how small the negative part of the quasiprobability
distribution can be made: the larger this minimal negative mass the
more contextual the system. This measure is not naturally extendable
to a measure of noncontextuality because all noncontextual systems
are identically characterized by this negative mass being zero. 

\subsection{}

We also consider the measure of contextuality called \emph{contextual
fraction}, proposed in Refs. \cite{AbramskyBrand2011,Amselem2012,ElitzurPopescuRohrlich1992}
and developed in Ref. \cite{AbramBarbMans2017}. The logic of this
measure is similar to that of $\textnormal{CNT}_{3}$: contextuality
is measured by how close certain quasiprobabilities (in this case,
nonnegative numbers allowed to sum to less than unity) can be made
to a proper probability distribution. The measure has been only formulated
under the constraint that random variables measuring the same property
in different contexts are identically distributed. This constraint,
called \emph{consistent connectedness} in CbD, is more generally known
as the \emph{no-disturbance} principle (or ``no-signaling'', in
the case of spatially distributed systems). We provide a CbD-based
generalization of contextual fraction to arbitrary systems, and show
that this measure, too, does not have a natural noncontextuality counterpart.

\subsection{}

It is important here to dispel a possible confusion. Any measure of
the degree of contextuality in a contextual system can be associated
with some complementary measure that can be interpreted as the degree
of noncontextuality in this contextual system. Thus, in the opening
example of this paper, if $F\left(\mathcal{R}\right)>0$, one could
define its contextuality degree as $F\left(\mathcal{R}\right)/b$,
and consider $1-F\left(\mathcal{R}\right)/b$ the ``degree of noncontextuality''
in this contextual system. In Abramsky, Barbosa, and Mansfield's construction
\cite{AbramBarbMans2017} a generalized version of this construction
is introduced explicitly: 1 minus contextual fraction is \emph{noncontextual
fraction}, for any contextual system. Our usage of the term ``measure
of noncontextuality'' is different --- it refers to a \emph{degree
of noncontextuality in a noncontextual system}. The noncontextual
fraction of Abramsky and colleagues is not a measure of noncontextuality
in this sense because it identically equals 1 for all noncontextual
systems. Returning to our opening example, $1-F\left(\mathcal{R}\right)/b$
is not a measure of noncontextuality in our sense, because it is predicated
on contextuality, $F\left(\mathcal{R}\right)>0$. If $F\left(\mathcal{R}\right)\leq0$,
something like $F\left(\mathcal{R}\right)/a$, assuming $a<0$, would
be an appropriate measure of noncontextuality (and, if one so wishes,
one could define $1-F\left(\mathcal{R}\right)/a$ as the ``contextual
fraction'' of this noncontextual system). 

\subsection{}

One's interest to measures of noncontextuality can be justified in
essentially the same way as one's interest to measures of contextuality,
except that one knows much more about the latter. Thus, one should
be interested if a system, be it contextual or noncontextual, is stably
so: whether a small perturbation of the random variables it is comprised
of will change its (non)contextuality status. Larger values of (non)contextuality
mean more stable (non)contextuality. A closely related reason is statistical.
If contextuality or noncontextuality of a system is established on
a sample level, one should be interested in whether this finding is
reliable: e.g., whether a high-level confidence interval for its contextuality
or noncontextuality degree lies entirely in the range of contextuality
or noncontextuality values, respectively. This is an especially important
task in fields outside quantum physics, e.g., in the contextuality
analysis of human behavior \cite{CervDzhSQ,DzhZhaKuj2016,BasievaJEP:G}.
Non-physical applications provide additional reasons for one's interest
in measures of noncontextuality: e.g., some models of decision making
can predict both contextual and noncontextual systems of random variables,
and noncontextual systems may be linked to features of decision making
that are, if anything, of greater interest than those in contextual
systems \cite{BasievaJEP:G}. In quantum physics, there is a growing
interest to the question of whether certain classical systems could
exhibit contextuality similar to that found in quantum systems \cite{Zhangetal.2019,Markiewiczetal.2019,Frustagliaetal2016}.
Here, it might be useful to quantify the ``classicality'' and ``non-classicality''
of systems by measures of, respectively, noncontextuality and contextuality,
preferably chosen so that they form each other's natural extensions.
One can argue that degree of contextuality has been linked to quantum
advantage in computation, communication complexity, and other matters
of intrinsic or practical interest \cite{Bermejoetal2017,Howardetal.2014,Brukneretal.2004},
while nothing like this is currently known about noncontextuality.
However, this may very well be due to the simple fact that no measures
of noncontextuality have so far been proposed and studied. 

\subsection{}

The three measures of contextuality considered in this paper are CbD-based,
which means that they are not constrained by the assumption of consistent
connectedness. We had to leave out a large number of interesting contextuality
measures discussed in the literature under the consistent connectedness
constraint (with the exception of the contextual fraction measure
that we generalize to apply to arbitrary systems). Thus, most of the
measures of nonlocality (as a special case of contextuality) reviewed
in Ref. \cite{Brunneretal2014} cannot be naturally extended to measures
of locality (noncontextuality). The measures of contextuality constructed
in relation to Bell-type criteria of noncontextuality for consistently
connected systems, as, e.g., in Ref. \cite{AmaralCunhaCabello2015},
usually can be extended to measures of noncontextuality along the
lines of our opening example. The fact that we do not discuss these
measures in detail is a reflection of the focus of this paper rather
than our view of their relative importance.

\section{Basics of the Contextuality-by-Default approach}

\subsection{}

A \emph{system} $\mathcal{R}$ of random variables is a set whose
elements are random variables $R_{q}^{c}$ labeled in two ways: by
their \emph{contents} $q\in Q$ (that which the random variable measures
or responds to) and their \emph{contexts} $c\in C$ (the conditions
under which this random variable is recorded): 
\begin{equation}
\mathcal{R}=\left\{ R_{q}^{c}:c\in C,q\in Q,q\prec c\right\} ,
\end{equation}
where $q\prec c$ indicates that content $q$ is measured (or responded
to) in context $c$. Throughout this paper, the set of contents $Q$
and the set of contexts $C$ are finite, and all random variables
in the system are Bernoulli, with values $0/1$.

\subsection{}

In CbD, with no loss of generality, one can make $q\prec c$ hold
true for all $q$ and $c$, by placing in every ``empty'' $\left(q,c\right)$-cell
a dummy variable with a single possible value \cite{Dzh2017Nothing}.
We will not be using this construction in this paper, as it is convenient
to think of the relation $\prec$ as the \emph{format} of the system
$\mathcal{R}$, the arrangement of the random variables without information
of their distributions.

\subsection{}

In each context $c$, the subset of random variables 
\begin{equation}
R^{c}=\left\{ R_{q}^{c}:q\in Q,q\prec c\right\} 
\end{equation}
is \emph{jointly distributed}, i.e., it is a random variable in its
own right. It is referred to as the \emph{bunch} for (or corresponding
to) context $c$. For each content $q$, the subset of random variables
\begin{equation}
\mathcal{R}_{q}=\left\{ R_{q}^{c}:c\in C,q\prec c\right\} 
\end{equation}
is referred to as the \emph{connection} for (or corresponding to)
content $q$. The elements of a connection are not jointly distributed,
they are \emph{stochastically unrelated}. (This is reflected in the
notation, $R$ vs $\mathcal{R}$: the bunch $R^{c}$ is a random variable
in its own right, while the connection $\mathcal{R}_{q}$ is not.)
More generally, any $R_{q}^{c}$ and $R_{q'}^{c'}$ are stochastically
unrelated unless $c=c'$. The terminology above is illustrated in
Fig. \ref{fig: system=00003Dformat}.\footnote{One might protest that given the notion of a context and a content,
the corresponding notions of a bunch and a connection are unnecessary.
It is indeed possible, as we have done in some of our publications,
to avoid the use of the latter two terms by speaking instead of context-sharing
and content-sharing variables. However, in discussing measures of
(non)contextuality and algorithms computing them, the use of the terms
in question, e.g., when speaking of ``bunch probabilities'' and
``connection probabilities'', is convenient.}

\begin{figure}
\[
(a)\quad\begin{array}{|c|c|c|c||c|}
\hline R_{1}^{1} & R_{2}^{1} &  &  & c^{1}\\
\hline  & R_{2}^{2} & R_{3}^{2} & R_{4}^{2} & c^{2}\\
\hline R_{1}^{3} &  & R_{3}^{3} &  & c^{3}\\
\hline R_{1}^{4} &  &  & R_{4}^{4} & c^{4}\\
\hline R_{1}^{5} & R_{2}^{5} & R_{3}^{5} &  & c^{5}\\
\hline\hline q_{1} & q_{2} & q_{3} & q_{4} & \mathcal{R}
\\\hline \end{array}\quad\quad\begin{array}{|c|c|c|c||c|}
\hline \star & \star &  &  & c^{1}\\
\hline  & \star & \star & \star & c^{2}\\
\hline \star &  & \star &  & c^{3}\\
\hline \star &  &  & \star & c^{4}\\
\hline \star & \star & \star &  & c^{5}\\
\hline\hline q_{1} & q_{2} & q_{3} & q_{4} & \mathcal{R}
\\\hline \end{array}\quad(b)
\]

\medskip{}

\[
(c)\quad\begin{array}{ccccc||c|}
\hline R^{1} & \star & \star &  &  & c^{1}\\
\hline R^{2} &  & \star & \star & \star & c^{2}\\
\hline R^{3} & \star &  & \star &  & c^{3}\\
\hline R^{4} & \star &  &  & \star & c^{4}\\
\hline R^{5} & \star & \star & \star &  & c^{5}\\
\hline\hline  & q_{1} & q_{2} & q_{3} & q_{4} & \mathcal{R}
\\\hline \end{array}\quad\quad\begin{array}{|c|c|c|c||c|}
\mathcal{R}_{1} & \mathcal{R}_{2} & \mathcal{R}_{3} & \mathcal{R}_{4} & \\
\star & \star &  &  & c^{1}\\
 & \star & \star & \star & c^{2}\\
\star &  & \star &  & c^{3}\\
\star &  &  & \star & c^{4}\\
\star & \star & \star &  & c^{5}\\
\hline\hline q_{1} & q_{2} & q_{3} & q_{4} & \mathcal{R}
\\\hline \end{array}\quad(d)
\]

\caption{Illustration for the basic terms of CbD. (a) A system $\mathcal{R}$
of random variables, with 4 contents measured in 5 contexts. Each
variable in the system is uniquely identified by its content and its
context. All random variables are dichotomous, $0/1$. (b) The format
of the system $\mathcal{R}$, showing which content is measured in
which context. It can also be used as a simplified representation
of $\mathcal{R}$, since the identification $R_{q}^{c}$ is uniquely
reconstructed from the position of the corresponding star. (c) Shows
5 bunches of the system. The random variables within a bunch are jointly
distributed, i.e., each bunch is a random variable. (d) Shows 4 connections
of the system. The random variables within a connection are stochastically
unrelated.\label{fig: system=00003Dformat}}
\end{figure}

\subsection{}

A system is \emph{consistently connected} (satisfies the ``no-disturbance''
requirement) if the distribution of each random variable in it depends
on its content only. If this is not the case, the system is \emph{inconsistently
connected}. The latter term can also be used for arbitrary systems,
that may but need not be consistently connected. Consistent connectedness
(non-disturbance) can sometimes be understood in the \emph{strong
sense,} as in Refs. \cite{AbramskyBrand2011,Dzh2003}: if several
contents $q_{1},\ldots,q_{k}$ are measured in two contents $c,c'$,
then the joint distributions of $\left\{ R_{q_{1}}^{c},\ldots,R_{q_{k}}^{c}\right\} $
and $\left\{ R_{q_{1}}^{c'},\ldots,R_{q_{k}}^{c'}\right\} $ coincide.
Nothing will change in this paper if consistent connectedness is understood
in this strong sense, because we generally assume systems are not
consistently connected even in the weaker, more general sense.

\subsection{}

The general definition of a \emph{coupling} for an indexed set $\mathcal{X}$
of random variables is that it is a jointly distributed and identically
indexed set $Y$ of random variables such that, for any subset $X$
of $\mathcal{X}$ possessing a joint distribution, the corresponding
subset of $Y$ is identically distributed. In particular, every element
of $\mathcal{X}$ is distributed as the corresponding element of $Y$.
In accordance with this general definition, a coupling of a system
$\mathcal{R}$ is a set of jointly distributed random variables 
\begin{equation}
S=\left\{ S_{q}^{c}:c\in C,q\in Q,q\prec c\right\} ,
\end{equation}
such that, for any context $c$, the bunches $S^{c}$ and $R^{c}$
are identically distributed.

\subsection{}

\label{subsec coupling}In CbD, we are interested not in just any
coupling of $\mathcal{R}$ but in those with a certain property. It
can be introduced as follows. For each separately taken connection
$\mathcal{R}_{q}$ one can find its \emph{multimaximal coupling} 
\begin{equation}
T_{q}=\left\{ T_{q}^{c}:c\in C,q\prec c\right\} ,
\end{equation}
defined as a coupling in which, for every $c,c'$ in which $q$ is
measured, the probability of $T_{q}^{c}=T_{q}^{c'}$ is maximal possible
(among all possible couplings of $\mathcal{R}_{q}$, or, equivalently,
for given marginal distributions of $R_{q}^{c}$ and $R_{q}^{c'}$).
With only dichotomous variables in play, every connection has a unique
multimaximal coupling, and for any subset $\left\{ R_{q}^{c},R_{q}^{c'},\ldots,R_{q}^{c''\ldots'}\right\} $
of a connection $\mathcal{R}_{q}$, the probability of $T_{q}^{c}=T_{q}^{c'}=\ldots=T_{q}^{c''\ldots'}$
is maximal possible \cite{DzhKuj2017.2.0,DzhKuj2017Fortsch}. It should
also be noted that maximization of $T_{q}^{c}=T_{q}^{c'}=\ldots=T_{q}^{c''\ldots'}$
means that both 
\begin{equation}
T_{q}^{c}=T_{q}^{c'}=\ldots=T_{q}^{c''\ldots'}=1
\end{equation}
and
\begin{equation}
T_{q}^{c}=T_{q}^{c'}=\ldots=T_{q}^{c''\ldots'}=0
\end{equation}
are maximized (recall that the values of all variables are encoded
$0/1$).

\subsection{}

A system $\mathcal{R}$ is defined as \emph{noncontextual} if it has
a coupling $S$ whose restrictions to all connections are multimaximal
couplings of these connections. The system is \emph{contextual} if
no such coupling $S$ exists. Equivalently, a system $\mathcal{R}$
is noncontextual if it has a coupling $S$ which is also a coupling
for the multimaximal couplings $T_{q}$ of connections $\mathcal{R}_{q}$
($q\in Q$). Yet another way of saying this is to define $\mathcal{R}$
as contextual if its bunches are incompatible with the multimaximal
couplings of its connections (cannot be ``sewn together'' within
a single overall distribution).

\section{Vectorial representation of systems}

\subsection{}

Any system $\mathcal{R}$ can be described by a \emph{vector of bunch
probabilities}. Abramsky and Brandenburger \cite{AbramskyBrand2011}
call it an \emph{empirical model}. For any context $c$, assuming
the $n_{c}$ random variables in its bunch were enumerated $R_{1}^{c},\ldots,R_{n_{c}}^{c}$,
we define 
\begin{equation}
\mathbf{p}^{\left(c\right)}=\left[\begin{array}{c}
p_{1}^{\left(c\right)}\\
\vdots\\
p_{i}^{\left(c\right)}\\
\vdots\\
p_{2^{n_{c}}}^{\left(c\right)}
\end{array}\right]=\left[\begin{array}{c}
\Pr\left[R_{1}^{c}=0,\ldots,R_{j}^{c}=0,\ldots,R_{n_{c}}^{c}=0\right]\\
\vdots\\
\Pr\left[R_{1}^{c}=r_{1},\ldots,R_{j}^{c}=r_{j},\ldots,R_{n_{c}}^{c}=r_{n_{c}}\right]\\
\vdots\\
\Pr\left[R_{1}^{c}=1,\ldots,R_{j}^{c}=1,\ldots,R_{n_{c}}^{c}=1\right]
\end{array}\right],\label{eq: bunch probs}
\end{equation}
where $\Pr$ stands for probability, and $r_{1},\ldots,r_{n_{c}}$
run through all $2^{n_{c}}$ combinations of $0/1$'s. The vector
of bunch probabilities is defined as 
\begin{equation}
\mathbf{\mathbf{p_{\left(b\right)}}}=\left[\begin{array}{c}
\mathbf{p}^{c_{1}}\\
\vdots\\
\mathbf{p}^{c_{j}}\\
\vdots\\
\mathbf{p}^{c_{|C|}}
\end{array}\right]
\end{equation}
(the boldface index $\mathbf{b}$ stands for ``bunches'').

\subsection{}

For any content $q,$ assuming the $m_{q}$ elements of the corresponding
connection were enumerated $R_{q}^{1},\ldots,R_{q}^{m_{q}}$, any
coupling $\left(T_{q}^{1},\ldots,T_{q}^{m_{q}}\right)$ (not necessarily
multimaximal) of this connection is defined by

\begin{equation}
\mathbf{p}_{\left(q\right)}=\left[\begin{array}{c}
p_{\left(q\right),1}\\
\vdots\\
p_{\left(q\right),i}\\
\vdots\\
p_{\left(q\right),2^{m_{q}}}
\end{array}\right]=\left[\begin{array}{c}
\Pr\left[T_{q}^{1}=0,\ldots,T_{q}^{j}=0,\ldots,T_{q}^{m_{q}}=0\right]\\
\vdots\\
\Pr\left[T_{q}^{1}=s_{1},\ldots,T_{q}^{j}=s_{j},\ldots,T_{q}^{m_{q}}=s_{m_{q}}\right]\\
\vdots\\
\Pr\left[T_{q}^{1}=1,\ldots,T_{q}^{j}=1,\ldots,T_{q}^{m_{q}}=1\right]
\end{array}\right],\label{eq: connection probs}
\end{equation}
with the same meaning of the terms as in (\ref{eq: bunch probs}).
The \emph{vector of connection probabilities} is defined as 
\begin{equation}
\mathbf{p_{\left(c\right)}}=\left[\begin{array}{c}
\mathbf{p}_{q_{1}}\\
\vdots\\
\mathbf{p}_{q_{j}}\\
\vdots\\
\mathbf{p}_{q_{|Q|}}
\end{array}\right]
\end{equation}
(the boldface index $\mathbf{c}$ stands for ``connections'').

\subsection{}

Finally we stack up the two vectors, for bunches and for connections,
to obtain the \emph{complete vector of probabilities}. 
\begin{equation}
\mathbf{p_{\left(\cdot\right)}}=\left[\begin{array}{c}
\mathbf{p_{\left(b\right)}}\\
\mathbf{p_{\left(c\right)}}
\end{array}\right].\label{eq: complete p}
\end{equation}
We include connection couplings in this representation of the system
even though they are computed rather than observed. This means that
one and the same system can be represented by multiple probability
vectors, depending on the couplings we choose for connections.

\subsection{}

Complete vectors of probabilities will be used in Section \ref{sec: Contextual-fraction},
when we discuss a generalized version of a measure proposed in Ref.
\cite{AbramBarbMans2017}. However, for purposes of computing CbD-based
measures of contextuality, $\mathbf{p_{\left(\cdot\right)}}$ is not
convenient because of its redundancy: one cannot change any component
of $\mathbf{p}^{\left(c\right)}$ or $\mathbf{p}_{\left(q\right)}$
without changing some of its other components. We will deal therefore
with one of the numerous versions of a \emph{reduced vector of probabilities}
in which components can be changed independently. The variant we choose
is introduced in Ref. \cite{DzhKuj2016}. It is based on the idea
of replacing $\mathbf{p}^{\left(c\right)}$ in (\ref{eq: bunch probs})
with probabilities 
\begin{equation}
\Pr\left[R_{q_{i}}^{c}=1:i\in I\right]=\left\langle \prod_{i\in I}R_{q_{i}}^{c}\right\rangle 
\end{equation}
for various subsets $I$ of $\left\{ 1,\ldots,n_{c}\right\} $. Analogously,
$\mathbf{p}_{\left(q\right)}$ in (\ref{eq: connection probs}) is
replaced with probabilities 
\begin{equation}
\Pr\left[T_{q}^{c_{j}}=1:j\in J\right]=\left\langle \prod_{j\in J}T_{q}^{c_{j}}\right\rangle 
\end{equation}
for various subsets $J$ of $\left\{ 1,\ldots,m_{q}\right\} $. These
probabilities (and also the events whose probabilities they are, when
this cannot cause confusion) are referred to as \emph{$k$-marginals},
where $k=0,1,2,\ldots$ is the \emph{order of the marginals} (the
number of the random variables involved). The 0-marginal is a constant
$\left\langle \right\rangle $ taken to be 1, and the 1-marginals
\begin{equation}
\left\langle R_{q}^{c}\right\rangle =\Pr\left[R_{q}^{c}=1\right]=\Pr\left[T_{q}^{c}=1\right]=\left\langle T_{q}^{c}\right\rangle 
\end{equation}
are shared by the bunches and the connections. Because of this, to
avoid redundancy, we put the 0-marginal and all 1-marginals in one
group, all higher-order marginals for bunches into a second group,
and all higher-order marginals for (couplings of) connections into
a third group.

\subsection{}

Let us order in some way all random variables in the system: $R_{q_{1}}^{c_{1}},\ldots,R_{q_{N}}^{c_{N}}$
. Define 
\begin{equation}
\mathbf{p}_{\mathbf{l}}=\left[\begin{array}{c}
1\\
\Pr\left[R_{q_{1}}^{c_{1}}=1\right]\\
\vdots\\
\Pr\left[R_{q_{i}}^{c_{i}}=1\right]\\
\vdots\\
\Pr\left[R_{q_{N}}^{c_{N}}=1\right]
\end{array}\right]=\left[\begin{array}{c}
\left\langle \right\rangle \\
\left\langle R_{q_{1}}^{c_{1}}\right\rangle \\
\vdots\\
\left\langle R_{q_{i}}^{c_{i}}\right\rangle \\
\vdots\\
\left\langle R_{q_{N}}^{c_{N}}\right\rangle 
\end{array}\right],
\end{equation}
where the boldface index $\mathbf{l}$ stands for ``low-order marginals''.

\subsection{}

For a given context $c$, let us enumerate $1,\ldots,2^{n_{c}}-n_{c}-1$
all nonempty and non-singleton subsets of the corresponding bunch:
$\binom{n_{c}}{2}$ 2-marginals followed by $\binom{n_{c}}{3}$ 3-marginals
etc. Define 
\begin{equation}
\mathbf{p}^{c}=\left[\begin{array}{c}
p_{1}^{c}\\
\vdots\\
p_{i}^{c}\\
\vdots\\
p_{2^{n_{c}}-n_{c}-1}^{c}
\end{array}\right]=\left[\begin{array}{c}
\Pr\left[R_{1}^{c}=1,R_{2}^{c}=1\right]\\
\vdots\\
\Pr\left[R_{i,1}^{c}=1,\ldots,R_{i,j}^{c}=1,\ldots,R_{i,n_{i,c}}^{c}=1\right]\\
\vdots\\
\Pr\left[R_{1}^{c}=1,\ldots,R_{j}^{c}=1,\ldots,R_{n_{c}}^{c}=1\right]
\end{array}\right]=\left[\begin{array}{c}
\left\langle R_{1}^{c}R_{2}^{c}\right\rangle \\
\vdots\\
\left\langle R_{i,1}^{c}\ldots R_{i,j}^{c}\ldots R_{i,n_{i,c}}^{c}\right\rangle \\
\vdots\\
\left\langle R_{1}^{c}\ldots R_{j}^{c}\ldots R_{n_{c}}^{c}\right\rangle 
\end{array}\right],\label{eq: reduced bunch}
\end{equation}
and the \emph{reduced vector of bunch probabilities} 
\begin{equation}
\mathbf{\mathbf{p_{b}}}=\left[\begin{array}{c}
\mathbf{p}^{c_{1}}\\
\vdots\\
\mathbf{p}^{c_{j}}\\
\vdots\\
\mathbf{p}^{c_{|C|}}
\end{array}\right].
\end{equation}

\subsection{}

We analogously define, having imposed some couplings (not necessarily
multimaximal) on the connections,

\begin{equation}
\mathbf{p}_{q}=\left[\begin{array}{c}
p_{q,1}\\
\vdots\\
p_{q,i}\\
\vdots\\
p_{q,2^{m_{q}}-m_{q}-1}
\end{array}\right]=\left[\begin{array}{c}
\Pr\left[T_{q}^{1}=1,T_{q}^{2}=1\right]\\
\vdots\\
\Pr\left[T_{q}^{i,1}=1,\ldots,T_{q}^{i,j}=1,\ldots,T_{q}^{i,m_{i,q}}=1\right]\\
\vdots\\
\Pr\left[T_{q}^{1}=1,\ldots,T_{q}^{j}=1,\ldots,T_{q}^{m_{q}}=1\right]
\end{array}\right]=\left[\begin{array}{c}
\left\langle T_{q}^{1}T_{q}^{2}\right\rangle \\
\vdots\\
\left\langle T_{q}^{i,1}\ldots T_{q}^{i,j}\ldots T_{q}^{i,m_{i,q}}\right\rangle \\
\vdots\\
\left\langle T_{q}^{1}\ldots T_{q}^{j}\ldots T_{q}^{m_{q}}\right\rangle 
\end{array}\right].\label{eq: reduced connection}
\end{equation}
The \emph{reduced vector of connection probabilities} is

\begin{equation}
\mathbf{\mathbf{p_{c}}}=\left[\begin{array}{c}
\mathbf{p}_{q_{1}}\\
\vdots\\
\mathbf{p}_{q_{j}}\\
\vdots\\
\mathbf{p}_{q_{|Q|}}
\end{array}\right].
\end{equation}
As mentioned in Section \ref{subsec coupling}, with reference to
\cite{DzhKuj2017.2.0,DzhKuj2017Fortsch}, if the couplings of connections
are chosen to be multimaximal, then all these probabilities are maximal
possible, given the values of the corresponding $1$-marginal probabilities.

\subsection{}

\label{subsec just pairs}Without loss of generality, one can delete
from $\mathbf{p_{c}}$ all $k$-marginals with $k>2$. As shown in
Ref. \cite{DzhKuj2017.2.0}, the 1-marginals and 2-marginals define
multimaximal couplings of the connections uniquely, and this makes
them sufficient for all CbD-based measures of contextuality.

\subsection{}

Finally, 
\begin{equation}
\mathbf{p}=\left[\begin{array}{c}
\mathbf{p_{l}}\\
\mathbf{p_{b}}\\
\mathbf{p_{c}}
\end{array}\right]
\end{equation}
is the \emph{reduced vector of probabilities} representing system\emph{
$\mathcal{R}$.} 

\subsection{}

Any component of $\mathbf{p}^{c}$ or of $\mathbf{p}_{q}$ can generally
change its value while other components remain fixed (which is impossible
in $\mathbf{p}^{\left(c\right)}$ and $\mathbf{p}_{\left(q\right)}$).
However, the range of possible changes is limited: every $k$-marginal
probability is limited from above by any $k-1$-marginal it contains,
and from below by any $\left(k+1\right)$-marginal containing it.

\section{\label{sec: Contextuality-in-vectorial}Contextuality in vectorial
representation}

\subsection{}

Consider a system $\mathcal{R}$ with $N$ dichotomous random variables,
and let $\mathbf{v}$ be the $2^{N}$-component vector of possible
values of a(ny) coupling $S$ of the entire system. An element of
$\mathbf{v}$ can be viewed as a conjunction of events 
\begin{equation}
\left\{ S_{q}^{c}=r_{q}^{c}:c\in C,q\in Q,q\prec c\right\} ,
\end{equation}
with $r_{q}^{c}=0/1$. Then any given $S$ is specified by a $2^{N}$-vector
$\mathbf{x}$ of the probabilities with which the corresponding elements
of $\mathbf{v}$ occur. Clearly, 
\begin{equation}
\mathbf{x}\geq0,\left\Vert \mathbf{x}\right\Vert =1,\label{eq: 2 constraints on X}
\end{equation}
where the inequality is componentwise, and the norm is $L_{1}$. We
call $\mathbf{x}$ a \emph{coupling vector} for $\mathcal{R}$.

\subsection{}

Let $\mathbf{p}$ be a (reduced) vector of probabilities. Then the
\emph{i}th component of $\mathbf{p}$ is the joint probability 
\begin{equation}
\Pr\left[S_{q}^{c}=1:\left(c,q\right)\in D_{i}\right]=\left\langle \prod_{\left(c,q\right)\in D_{i}}S_{q}^{c}\right\rangle \label{eq: event}
\end{equation}
for some $D_{i}$. The latter can be a low-marginal event, in which
case it is empty or a singleton; or $D_{i}$ can be a bunch event,
in which case it consists of a fixed $c$ paired with two or more
$q$'s; or else it can be a connection event, in which case it has
a fixed $q$ paired with two or more $c$'s (or with precisely two
$c$'s, in view of Section \ref{subsec just pairs}).

\subsection{}

We now construct a Boolean matrix $\mathbf{M}$ having $2^{N}$ columns,
with the $j$th column being labeled by the $j$th value of $\mathbf{v}$
($j=1,\ldots,2^{N}$). This matrix is the same for all systems in
the format of $\mathcal{R}$. The $i$th row of this matrix is labeled
by the event (\ref{eq: event}) whose probability is the $i$th element
of $\mathbf{p}$. If all the random variables in this event equal
to 1 in the $j$th value of $\mathbf{v}$, then we put 1 in the cell
$\left(i,j\right)$ of $\mathbf{M}$. All other cells of \textbf{$\mathbf{M}$}
are filled with zeros.

\subsection{}

The matrix can be presented as

\begin{equation}
\mathbf{M}=\left(\begin{array}{c}
\mathbf{M_{l}}\\
\mathbf{M_{b}}\\
\mathbf{M_{c}}
\end{array}\right),\label{eq: M}
\end{equation}
with $\mathbf{M_{l}}$, $\mathbf{M_{b}}$, and $\mathbf{M_{c}}$ corresponding
to the $\mathbf{p_{l}}$-part (low-marginal probabilities), $\mathbf{p_{b}}$-part
(bunch probabilities), and $\mathbf{p_{c}}$-part (connection probabilities)
of $\mathbf{p}$, respectively. In particular, the first row of $\mathbf{M}$
corresponds to the zero-marginal 1, and this row contains 1 in all
cells.

\subsection{}

Let 
\begin{equation}
\mathbf{p^{*}=}\left[\begin{array}{c}
\mathbf{p_{l}^{*}}\\
\mathbf{p_{b}^{*}}\\
\mathbf{p_{c}^{*}}
\end{array}\right]
\end{equation}
be a (reduced) vector of probabilities whose $\mathbf{p_{l}}$-part
and $\mathbf{p_{b}}$-part consist of \emph{empirical probabilities}
(estimated from an experiment or predicted by a model), and $\mathbf{p_{c}}$-part
consists of the connection probabilities for multimaximal couplings.
Then the system $\mathcal{R}$ represented by $\mathbf{p^{*}}$ is
noncontextual if and only if 
\begin{equation}
\mathbf{M}\mathbf{x}=\mathbf{p}^{*}
\end{equation}
for some nonnegative coupling vector $\mathbf{x}$. If no such nonnegative
$\mathbf{x}$ exists, then $\mathcal{R}$ is contextual. In reference
to (\ref{eq: 2 constraints on X}), note that $\left\Vert \mathbf{x}\right\Vert =1$
is guaranteed by the first row of $\mathbf{M}$ (consisting of $1$'s
only) and first element of $\mathbf{p}^{*}$ ($\left\langle \right\rangle =1$).

\subsection{}

As a step towards measures of contextuality, consider the convex polytope

\begin{equation}
\mathbb{P}=\left\{ \mathbf{p}:\mathbf{M}\mathbf{x}=\mathbf{p},\textnormal{ for some }\mathbf{x}\geq0\right\} .
\end{equation}
It corresponds to the set of all possible couplings of all systems
having the same format as $\mathcal{R}$ (because matrices $\mathbf{M}$
are in a one-to-one correspondence with system formats).

\subsection{}

A specific system $\mathcal{R}$ is defined by specifying the vectors
$\mathbf{p_{l}}=\mathbf{p_{l}^{*}}$ and $\mathbf{p_{b}}=\mathbf{p_{b}^{*}}$.
This defines a convex polytope which is a cross-section of the polytope
$\mathbb{P}$, 
\begin{equation}
\mathbb{P}_{\mathbf{c}}=\left\{ \mathbf{p_{c}}:\mathbf{M_{c}}\mathbf{x}=\mathbf{p_{c}},\textnormal{ for some }\mathbf{x}\geq0,\mathbf{M_{l}}\mathbf{x}=\mathbf{p_{l}^{*}},\mathbf{M_{b}}\mathbf{x}=\mathbf{p_{b}^{*}}\right\} .\label{eq: feasibillity}
\end{equation}
We refer to it as the \emph{feasibility polytope} (for system $\mathcal{R}$).
It corresponds to the set of all possible couplings of system $\mathcal{R}$.

\subsection{}

A symmetrically opposite construction is the convex polytope\emph{
\begin{equation}
\mathbb{P}_{\mathbf{b}}=\left\{ \mathbf{p_{b}}:\mathbf{M_{b}}\mathbf{x}=\mathbf{p_{b}},\textnormal{ for some }\mathbf{x}\geq0,\mathbf{M_{l}}\mathbf{x}=\mathbf{p_{l}^{*}},\mathbf{M_{c}}\mathbf{x}=\mathbf{p_{c}^{*}}\right\} .\label{eq: noncontextuality}
\end{equation}
}We can call it the \emph{noncontextuality polytope} (for system $\mathcal{R}$),
as it corresponds to all noncontextual systems with the same 1-marginals
as $\mathcal{R}$. Another way of describing $\mathbb{P}_{\mathbf{b}}$,
to emphasize its symmetry with $\mathbb{P}_{\mathbf{c}}$, is that
$\mathbb{P}_{\mathbf{b}}$ corresponds to all possible couplings of
the multimaximal couplings of the system's connections.

\subsection{}

Clearly, system $\mathcal{R}$ is noncontextual if and only if $\mathbf{p_{c}^{*}}\in\mathbb{P}_{\mathbf{c}}$
and $\mathbf{p_{b}^{*}}\in\mathbb{P}_{\mathbf{b}}$, with the two
statements implying each other,
\begin{equation}
\mathbf{p_{c}^{*}}\in\mathbb{P}_{\mathbf{c}}\Longleftrightarrow\mathbf{p_{b}^{*}}\in\mathbb{P}_{\mathbf{b}}.
\end{equation}

\section{Measures of contextuality}

\subsection{}

For a contextual system $\mathcal{R}$, $\mathbf{p_{c}^{*}}$ is outside
$\mathbb{P}_{\mathbf{c}}$, and the $L_{1}$-distance between them
is a natural measure of contextuality,

\begin{equation}
\textnormal{\ensuremath{\textnormal{CNT}_{1}}}=\min_{\mathbf{p_{c}}\in\mathbb{P}_{\mathbf{c}}}\left\Vert \mathbf{p_{c}^{*}}-\mathbf{p_{c}}\right\Vert =\mathbf{1}\cdot\mathbf{p_{c}^{*}}-\max_{\mathbf{p_{c}}\in\mathbb{P}_{\mathbf{c}}}\left(\mathbf{1}\cdot\mathbf{p_{c}}\right),
\end{equation}
where the equality follows from $\mathbf{p_{c}^{*}}\geq\mathbf{p_{c}}$
(componentwise). We can also write 
\begin{equation}
\textnormal{\ensuremath{\textnormal{CNT}_{1}}}=\mathbf{1}\cdot\mathbf{p_{c}^{*}}-\max_{\mathbf{x}\geq0,\mathbf{M_{l}}\mathbf{x}=\mathbf{p_{l}^{*}},\mathbf{M_{b}}\mathbf{x}=\mathbf{p_{b}^{*}}}\left(\mathbf{1}\cdot\mathbf{M_{c}}\mathbf{x}\right).
\end{equation}
To interpret, since the system $\mathcal{R}$ is contextual, its bunches
are incompatible with the multimaximal couplings of its connections.
$\ensuremath{\textnormal{CNT}_{1}}$ measures how close the couplings
of these connections that are compatible with the system's bunches
can be made to the multimaximal ones. 

\subsection{}

This measure was the first one proposed within the framework of CbD
\cite{KujDzhProof2016,KujDzhLar2015}.\footnote{\label{fn: More-precisely}More precisely, the measure proposed in
Refs. \cite{KujDzhProof2016,KujDzhLar2015} is twice $\ensuremath{\textnormal{CNT}_{1}}$.} Its linear programming implementation is 
\begin{equation}
\begin{array}{|ccc|}
\hline \textnormal{find } & \textnormal{maximizing} & \textnormal{subject to}\\
\mathbf{\mathbf{x}} & \mathbf{1}\cdot\mathbf{\mathbf{M_{c}}\mathbf{x}} & \mathbf{x\geq}0\\
 &  & \mathbf{M_{l}}\mathbf{x}=\mathbf{p_{l}^{*}}\\
 &  & \mathbf{M_{b}x}=\mathbf{\mathbf{p_{b}^{*}}}
\\\hline \end{array}\:.
\end{equation}
A solution $\mathbf{x}^{*}$ must exist, and any such $\mathbf{x}^{*}$
yields 
\begin{equation}
\textnormal{\ensuremath{\textnormal{CNT}_{1}}}=\mathbf{1}\cdot\left(\mathbf{p_{c}^{*}}-\mathbf{\mathbf{M_{c}}\mathbf{x^{*}}}\right)=\mathbf{1}\cdot\left(\mathbf{p}^{*}-\mathbf{\mathbf{\mathbf{M}\mathbf{x}^{*}}}\right).
\end{equation}

\subsection{}

The ``symmetrically opposite'' measure is the $L_{1}$-distance
of $\mathbf{p_{b}^{*}}$ from the contextuality polytope $\mathbb{P}_{\mathbf{b}}$,

\begin{equation}
\textnormal{\ensuremath{\textnormal{CNT}_{2}}}=\min_{\mathbf{p_{b}}\in\mathbb{P}_{\mathbf{b}}}\left\Vert \mathbf{p_{b}^{*}}-\mathbf{p_{b}}\right\Vert =\min_{\mathbf{x}\geq0,\mathbf{M_{l}}\mathbf{x}=\mathbf{p_{l}^{*}},\mathbf{M_{c}}\mathbf{x}=\mathbf{p_{c}^{*}}}\left\Vert \mathbf{p_{b}^{*}}-\mathbf{\mathbf{M_{b}}\mathbf{x}}\right\Vert .
\end{equation}
The interpretation should be clear. Since $\mathcal{R}$ is contextual,
the multimaximal couplings of its connections are not compatible with
its bunches. $\ensuremath{\textnormal{CNT}_{2}}$ measures how close
the bunches that are compatible with these multimaximal couplings
can be made to the observed bunches. 

\subsection{}

This measure is introduced here for the first time. Its linear programming
implementation is

\begin{equation}
\begin{array}{|ccc|}
\hline \textnormal{find } & \textnormal{minimizing} & \textnormal{subject to}\\
\mathbf{\mathbf{x}} & \mathbf{\mathbf{1}\cdot\mathbf{d}} & -\mathbf{d}\leq\mathbf{p_{b}^{*}}-\mathbf{\mathbf{M_{b}}\mathbf{x}}\leq\mathbf{d}\\
 &  & \mathbf{x,d\geq}0\\
 &  & \mathbf{M_{l}}\mathbf{x}=\mathbf{p_{l}^{*}}\\
 &  & \mathbf{M_{c}x}=\mathbf{\mathbf{p_{c}^{*}}}
\\\hline \end{array}\:.
\end{equation}
Again, for any solution $\mathbf{x}^{*}$, 
\begin{equation}
\textnormal{\ensuremath{\textnormal{CNT}_{2}}}=\left\Vert \mathbf{p_{b}}^{*}-\mathbf{\mathbf{M_{b}}\mathbf{x^{*}}}\right\Vert =\left\Vert \mathbf{p}^{*}-\mathbf{\mathbf{\mathbf{M}\mathbf{x}^{*}}}\right\Vert .
\end{equation}

\subsection{}

The third measure to consider, $\textnormal{CNT}_{3}$, has been proposed
in Ref. \cite{DzhKuj2016}, and brought to its present form in Refs.
\cite{DzhKuj2017.2.0,DzhCerKuj2017}. However, the ideas on which
it is based date back to Paul Dirac, with contemporary elaborations,
including relating it to contextuality, found in Refs. \cite{AbramskyBrand2011,Al-Safi2013,deBarrosOasSuppes2015,Spekkens20084,deBarrosDzhKujOas2016}.
The measure is based on the observation \cite{DzhKuj2016} that if
one drops the nonnegativity constraint in (\ref{eq: 2 constraints on X}),
replacing thereby probability distributions with signed-measure distributions,
then the set 
\begin{equation}
\mathbb{Y}=\left\{ \mathbf{y}:\mathbf{M}\mathbf{y}=\mathbf{p^{*}}\right\} 
\end{equation}
is nonempty. Clearly, 
\begin{equation}
\mathbf{y}\in\mathbb{Y}\Longrightarrow\mathbf{1}\cdot\left|\mathbf{y}\right|\geq1,
\end{equation}
were the absolute value is computed componentwise. The system then
is noncontextual if and only if there is a $\mathbf{y}\in\mathbb{Y}$
with 
\begin{equation}
\mathbf{1}\cdot\left|\mathbf{y}\right|=\mathbf{1}\cdot\mathbf{y}=1.
\end{equation}
It follows that

\begin{equation}
\textnormal{\ensuremath{\textnormal{CNT}_{3}}}=\min_{\mathbf{y\in\mathbb{Y}}}\left(\mathbf{1}\cdot\left|\mathbf{y}\right|\right)-1
\end{equation}
is a natural measure of contextuality. As shown in Ref. \cite{DzhKuj2016},
this minimum is always attained. The quantity $\mathbf{1}\cdot\left|\mathbf{y}\right|$
is called \emph{total variation} (of the signed measure), so $\textnormal{CNT}_{3}$
can be referred as the minimum total variation measure (minus 1). 

\subsection{}

The linear programming implementation of $\textnormal{CNT}_{3}$ is

\begin{equation}
\begin{array}{|ccc|}
\hline \textnormal{find } & \textnormal{minimizing} & \textnormal{subject to:}\\
\mathbf{\mathbf{y}_{+}},\mathbf{\mathbf{y}_{-}} & \mathbf{\mathbf{1}\cdot}\mathbf{\mathbf{\mathbf{y}_{-}}} & \mathbf{M\left(\mathbf{\mathbf{y}_{+}}\mathbf{\mathbf{-\mathbf{y}_{-}}}\right)}=\mathbf{p^{*}}\\
 &  & \mathbf{\mathbf{y}_{+}},\mathbf{\mathbf{\mathbf{y}_{-}}}\geq0
\\\hline \end{array}\:.
\end{equation}
With any solution $\mathbf{y}_{+}^{*},\mathbf{y}_{-}^{*}$, 
\begin{equation}
\textnormal{\ensuremath{\textnormal{CNT}_{3}}}=\mathbf{1}\cdot\left|\mathbf{y}_{+}^{*}-\mathbf{y}_{-}^{*}\right|-1.
\end{equation}

\section{Noncontextuality}

\subsection{}

Consider now the situation when $\mathcal{R}$ is noncontextual. With
respect to the $\textnormal{CNT}_{1}$ measure, this means that $\mathbf{p_{c}^{*}}\in\mathbb{P}_{\mathbf{c}}$.
The question we pose is whether $\textnormal{CNT}_{1}$ can be extended
into a noncontextuality measure by computing 
\begin{equation}
\ensuremath{\textnormal{NCNT}_{1}}=\inf_{\mathbf{p_{c}}\not\in\mathbb{P}_{\mathbf{c}}}\left\Vert \mathbf{p_{c}^{*}}-\mathbf{p_{c}}\right\Vert =\min_{\mathbf{p_{c}}\in\mathbb{\partial P}_{\mathbf{c}}}\left\Vert \mathbf{p_{c}^{*}}-\mathbf{p_{c}}\right\Vert ,
\end{equation}
where $\partial$(polytope) indicates the boundary of the polytope.
The answer to this question turns out to be negative: while this distance
is well-defined, it is zero for any $\mathbf{p_{c}^{*}}$. Indeed,
if $\mathbf{p_{c}^{*}}$ were an interior point of $\mathbb{P}_{\mathbf{c}}$,
one could increase some of the probabilities in (\ref{eq: reduced connection})
by a small amount and still remain within $\mathbb{P}_{\mathbf{c}}$.
But this is impossible, since all $k$-marginal probabilities with
$k>1$ have maximal possible values. $\textnormal{CNT}_{1}$ does
not have a noncontextual counterpart.

\subsection{}

The situation is different with $\textnormal{CNT}_{2}$. The measure
\begin{equation}
\textnormal{NCNT}_{2}=\inf_{\mathbf{p_{b}}\not\in\mathbb{P}_{\mathbf{b}}}\left\Vert \mathbf{p_{b}^{*}}-\mathbf{p_{b}}\right\Vert =\min_{\mathbf{p_{b}}\in\partial\mathbb{P}_{\mathbf{b}}}\left\Vert \mathbf{p_{b}^{*}}-\mathbf{p_{b}}\right\Vert 
\end{equation}
is well-defined and varies as $\mathbf{p_{b}^{*}}$ varies within
$\mathbb{P}_{\mathbf{b}}$. To interpret, since $\mathcal{R}$ is
noncontextual, the multimaximal couplings of its connections are compatible
with its bunches. $\textnormal{NCNT}_{2}$ measures how far these
bunches are from those that are not compatible with these multimaximal
couplings. To compute $\textnormal{NCNT}_{2}$ we can make use of
the following theorem \cite{Tuenter2006}: a point on the boundary
of a convex polytope $L_{1}$-closest to an interior point differs
from the latter in a single coordinate. This means that all we have
to do is to increase or decrease the probabilities in (\ref{eq: reduced bunch})
one by one as far as possible without leaving the polytope, and to
choose the smallest increase or decrease at the end.

\subsection{}

The linear programming implementation of this procedure is as follows.
Let all elements of $\mathbf{p_{b}^{*}}$ be enumerated $1,\ldots,K$.
Then, for every $i=1,\ldots,K$,

\begin{equation}
\begin{array}{|ccc|}
\hline \textnormal{find } & \textnormal{maximizing} & \textnormal{subject to}\\
d_{i}^{+},\mathbf{x} & d_{i}^{+} & \mathbf{p_{b}}^{*}+d_{i}^{+}\mathbf{e}_{i}=\mathbf{\mathbf{M_{b}}\mathbf{x}}\\
 &  & d_{i}^{+},\mathbf{x\geq}0\\
 &  & \mathbf{M_{l}}\mathbf{x}=\mathbf{p_{l}^{*}}\\
 &  & \mathbf{M_{c}x}=\mathbf{\mathbf{p_{c}^{*}}}
\\\hline \end{array}
\end{equation}
and 
\begin{equation}
\begin{array}{|ccc|}
\hline \textnormal{find } & \textnormal{maximizing} & \textnormal{subject to}\\
d_{i}^{-},\mathbf{x} & d_{i}^{-} & \mathbf{p_{b}}^{*}-d_{i}^{-}\mathbf{e}_{i}=\mathbf{\mathbf{M_{b}}\mathbf{x}}\\
 &  & d_{i}^{-},\mathbf{x\geq}0\\
 &  & \mathbf{M_{l}}\mathbf{x}=\mathbf{p_{l}^{*}}\\
 &  & \mathbf{M_{c}x}=\mathbf{\mathbf{p_{c}^{*}}}
\\\hline \end{array}\:,
\end{equation}
where $\mathbf{e}_{i}$ is the unit vector with the $i$th component
equal to 1. Once the solutions $d_{i}^{*+},d_{i}^{*-}$ for $i=1,\ldots,K$
are determined,

\begin{equation}
\textnormal{NCNT}_{2}=\min_{i=1,\ldots,K}\left\{ \min\left(d_{i}^{*+},d_{i}^{*-}\right)\right\} .
\end{equation}

\subsection{}

Consider now $\textnormal{CNT}_{3}$. If $\mathcal{R}$ is noncontextual,
there is a nonnegative $\mathbf{x}$ such that $\mathbf{M}\mathbf{x}=\mathbf{p^{*}}$,
and 
\[
\textnormal{\ensuremath{\textnormal{CNT}_{3}}}=\mathbf{1}\cdot\mathbf{x}-1=0.
\]
There seems to be no way to extend it to a noncontextuality measure
without modifying its logic.

\section{\label{sec: Contextual-fraction}Contextual fraction}

\subsection{}

We discuss next the measure called \emph{contextual fraction} (CNTF)
\cite{AbramBarbMans2017,AbramskyBrand2011} (see also Refs. \cite{ElitzurPopescuRohrlich1992,Amselem2012}).
It has been formulated for consistently connected systems only, and
its CbD-based generalization is not unique. We will consider one such
generalization, obtained by treating multimaximal couplings of connections
as if they were additional bunches. This can in fact be used to formally
redefine every system into a consistently connected one, as proposed
by Amaral, Duarte, and Oliveira in Ref. \cite{AmaralDuarteOliveira2018}.\footnote{Amaral and coauthors use maximal couplings rather than multimaximal
ones, and they allow for multivalued variables (as we did in the older
version of CbD, e.g., in Ref. \cite{DzhKuj2016}). The difference
between the two couplings of a set $\left\{ X_{1},\ldots,X_{n}\right\} $
is that in the multimaximal coupling $\left\{ Y_{1},\ldots,Y_{n}\right\} $
we maximize probabilities of all equalities $Y_{i}=Y_{j}$ (whence
it follows that we also maximize the probability of $Y_{i_{1}}=Y_{i_{2}}=\ldots=Y_{i_{k}}$
for any subset of $\left\{ Y_{1},\ldots,Y_{n}\right\} $), whereas
a maximal coupling $\left\{ Z_{1},\ldots,Z_{n}\right\} $ only maximizes
the single chain equality $Z_{1}=Z_{2}=\ldots=Z_{n}$. The reasons
we adhere to multimaximal couplings and canonical systems, with all
variables dichotomized, were laid out in Refs. \cite{DzhKuj2017.2.0,DzhKuj2017Fortsch,DzhCerKuj2017}.
Here, it will suffice to say that our multimaximal couplings are unique,
whereas maximal couplings generally are not, even for dichotomous
variables (if there are more than two of them). Since measures of
(non)contextuality generally depend on what couplings are being used,
the approach advocated in Ref. \cite{AmaralDuarteOliveira2018} faces
the problem of choice.} We prefer, however, to simply complement bunch probabilities with
connection probabilities rather than redefining the system itself.

\subsection{}

Unlike the three measures considered above, however, CNTF requires
that we deal with complete vectors of probability $\mathbf{p}_{\left(\cdot\right)}$,
as defined in (\ref{eq: complete p}), rather than reduced vectors.
We need to accordingly replace the Boolean matrix $\mathbf{M}$ with
the Boolean matrix $\mathbf{M}_{\left(\cdot\right)}$ such that the
system $\mathcal{R}$ represented by $\mathbf{p_{\left(\cdot\right)}^{*}}$
is noncontextual if and only if 
\begin{equation}
\mathbf{M}_{\left(\cdot\right)}\mathbf{x}=\mathbf{p_{\left(\cdot\right)}^{*}},
\end{equation}
with $\mathbf{x}$ defined as above. The structure of $\mathbf{M}_{\left(\cdot\right)}$
is described in Ref. \cite{DzhKuj2016}, and its summary is as follows.
Recall the definition of vector $\mathbf{v}$ in Section \ref{sec: Contextuality-in-vectorial}.
The columns of matrix $\mathbf{M}_{\left(\cdot\right)}$ are labeled
by the elements of $\mathbf{v}$ in the same way as in matrix $\mathbf{M}$.
The $i$th row of this matrix is labeled by the event whose probability
is the $i$th element of $\mathbf{p_{\left(\cdot\right)}^{*}}$ in
\ref{eq: bunch probs} and \ref{eq: connection probs}. If all the
random variables in this event have the same values in the $j$th
value of $\mathbf{v}$, then we put 1 in the cell $\left(i,j\right)$
of $\mathbf{M}_{\left(\cdot\right)}$. All other cells of \textbf{$\mathbf{M}_{\left(\cdot\right)}$}
are filled with zeros.

\subsection{}

Consider the convex polytope 
\begin{equation}
\mathbb{Z}=\{\mathbf{z}:\mathbf{M}_{\left(\cdot\right)}\mathbf{z}\leq\mathbf{\mathbf{p_{\left(\cdot\right)}^{*}}},\mathbf{z}\geq0,\mathbf{1}\cdot\mathbf{z}\leq1\}.\label{eq: CNTF_Z}
\end{equation}
This polytope is nonempty, because, e.g., it contains $\mathbf{z}=\mathbf{0}$.
If $\mathbb{Z}$ contains a $\mathbf{z}$ with $\mathbf{1}\cdot\mathbf{z}=1$,
then $\mathbf{M}_{\left(\cdot\right)}\mathbf{z}=\mathbf{\mathbf{p_{\left(\cdot\right)}^{*}}}$
because the elements of every $\mathbf{p}^{\left(c\right)}$ in (\ref{eq: bunch probs})
and every $\mathbf{p}_{\left(q\right)}$ and (\ref{eq: connection probs})
sum to 1, and the corresponding rows of $\mathbf{M}_{\left(\cdot\right)}$
sum to a row consisting of $1$'s only. Such a system therefore is
noncontextual. If $\mathbf{1}\cdot\mathbf{z}<1$ for all $\mathbf{z}\in\mathbb{Z}$,
the system is contextual, and its degree of contextuality can be measured
by the difference between $1$ and the maximal total mass $\mathbf{1}\cdot\mathbf{z}$
achievable in $\mathbb{Z}$: 
\begin{equation}
\textnormal{CNTF}=1-\max_{\mathbf{z}\in\mathbb{Z}}\left(\mathbf{1}\cdot\mathbf{z}\right).\label{eq: CNTF}
\end{equation}
The linear programming formulation of this measure is 
\begin{equation}
\begin{array}{|ccc|}
\hline \textnormal{find } & \textnormal{maximizing} & \textnormal{subject to:}\\
\mathbf{z} & \mathbf{\mathbf{1}\cdot}\mathbf{z} & \mathbf{M}_{\left(\cdot\right)}\mathbf{z}\leq\mathbf{\mathbf{p_{\left(\cdot\right)}^{*}}}\\
 &  & \mathbf{z\geq}0\\
 &  & \mathbf{1}\cdot\mathbf{z}\leq1
\\\hline \end{array}\:.\label{eq: CNTF_LP}
\end{equation}
Can this measure be naturally extended to a measure of noncontextuality?
The answer is negative, for the same reason as in the case of $\textnormal{CNT}_{3}$.
If $\mathbb{Z}$ contains a vector $\mathbf{z}$ such that $\mathbf{1}\cdot\mathbf{z}=1$,
then $\mathbf{M}_{\left(\cdot\right)}\mathbf{z}=\mathbf{\mathbf{p_{\left(\cdot\right)}^{*}}}$
and CNTF $=0$. We do not have conceptual means here to distinguish
different noncontextual systems.

\section{Conclusion}

\subsection{}

We have provided an overview of three CbD-based measures of contextuality.
Two of them, $\textnormal{CNT}_{1}$ and $\textnormal{CNT}_{2}$ are
$L_{1}$-distances between a probability vector representing a system
and a convex polytope. For either of the measures, if the probability
vector is not outside the polytope, a natural way of extending this
measure to a noncontextuality measure is to compute the $L_{1}$-distance
from the point to the surface of the polytope. We have seen, however,
that in the case of $\textnormal{CNT}_{1}$, the probability vector
never gets inside the feasibility polytope (\ref{eq: feasibillity}):
as $\textnormal{CNT}_{1}$ decreases to zero, and the vector becomes
noncontextual, it sticks to the polytope's surface. By contrast, for
$\textnormal{CNT}_{2}$, as its value decreases to zero, the vector
of probabilities continues to move inside the contextuality polytope
(\ref{eq: noncontextuality}). $\textnormal{CNT}_{2}$ therefore is
naturally extended to $\textnormal{NCNT}_{2}$ , the distance from
an interior point to the polytope's surface.

\subsection{}

The third CbD-based measure, $\textnormal{CNT}_{3}$, is of a different
kind. It measures the degree of contextuality by dropping the nonnegativity
constraint in (\ref{eq: 2 constraints on X}) for coupling vectors
$\mathbf{x}$, and determining how close the vectors $\mathbf{y}$
thus obtained can be made to a proper probability distribution while
satisfying $\mathbf{My=p^{*}}$. The logical structure of this measure
is close to Abramsky, Barbosa, and Mansfield's CNTF measure, CbD-generalized
to apply to arbitrary systems. Here the degree of contextuality is
measured by replacing the summing-to-unity constraint in (\ref{eq: 2 constraints on X})
with $\mathbf{\mathbf{1}\cdot z}\leq1$, and determining how close
the vectors $\mathbf{z}$ can be made to a proper probability distribution
while satisfying $\mathbf{M}_{\left(\cdot\right)}\mathbf{z}\leq\mathbf{\mathbf{p_{\left(\cdot\right)}^{*}}}$.
Both these measures do not lend themselves to natural noncontextuality
extensions. The values of these two measures, total variation less
1 in $\textnormal{CNT}_{3}$ and 1 minus total mass in CNTF, vary
with probability vectors representing contextual systems but freeze
at zero values for all noncontextual systems.

\subsection{}

The relationship between the four measures of contextuality discussed
in this paper is far from being clear. One nice feature of $\textnormal{CNT}_{1}$,
the oldest CbD-based measure, is that it is proportional to the violation
of the generalized Bell inequalities in the case of cyclic systems
\cite{KujDzhLar2015,DzhKujLar,KujDzhProof2016}: 
\begin{equation}
\textnormal{\ensuremath{\textnormal{CNT}_{1}}}=\frac{1}{4}\max_{\left(\iota_{1},\ldots,\iota_{k}\right)\in\left\{ -1,1\right\} ^{n}:\prod_{i=1}^{n}\iota_{i}=-1}\sum_{i=1}^{n}\iota_{i}\left\langle R_{i}^{i}R_{i\oplus1}^{i}\right\rangle -n+2-\sum_{i=1}^{n}\left|\left\langle R_{i}^{i}\right\rangle -\left\langle R_{i}^{i\ominus1}\right\rangle \right|,
\end{equation}
where the dichotomous variables are assumed to be $\pm1$-valued rather
than Bernoulli.\footnote{See footnote \ref{fn: More-precisely}.}
Here, the bunch for context $c_{i}$ ($i=1,\ldots,n$) consists of
two random variables $R_{i}^{i},R_{i\oplus1}^{i}$, where $i\oplus1=i+1$
for $i<n$ and $n\oplus1=1$. (This inequality generalizes to arbitrary
systems the inequality proved in Ref. \cite{Araujoetal2013}, in a
very different way, for consistently connected systems.) Our analysis
\cite{Dzh2019Cyclic} shows that in the case of cyclic systems 
\begin{equation}
\ensuremath{\textnormal{CNT}_{1}}=\ensuremath{\textnormal{CNT}_{2}.}
\end{equation}
It has been conjectured, based on numerical computations conducted
with the help of Víctor Cervantes, that in the case of cyclic systems
\begin{equation}
\mathsf{\textnormal{\ensuremath{\textnormal{CNT}_{3}}}}=\textnormal{2\ensuremath{\textnormal{CNT}_{1}}}/(n-1)=\textnormal{2\ensuremath{\textnormal{CNT}_{2}}}/(n-1).
\end{equation}
Beyond cyclic systems, however, we know that $\textnormal{CNT}_{1}$
and $\textnormal{CNT}_{3}$ are not generally related to each other
by any function \cite{deBarrosDzhKujOas2016}. The relations between
the three CbD-based measures and CNTF is yet to be investigated. 

\subsection{}

It is worth mentioning that all measures of contextuality (and noncontextuality)
involve a certain degree of arbitrariness. For instance, both $\textnormal{CNT}_{1}$
and $\textnormal{CNT}_{2}$ could be constructed with another $L_{p}$
or $L_{\infty}$ replacing $L_{1}$, and there seem to be no unchallengeable
principles to guide one's choice (although $L_{1}$ may be argued
to be preferable because of the additivity of probabilities). The
choice of a reduced vector of probabilities adds another dimension
of arbitrariness: although all reduced representations are linear
transformations of the complete one, $\min L_{1}$-distance values
for them may not be related to each other in a simple way. It seems
therefore that one could profitably use several measures to characterize
a given system. At the same time, the fact that only some measures
of contextuality naturally extend into measures of noncontextuality
may provide principled guidance in constraining the multitude of possibilities.

\paragraph{Acknowledgments.}

We are grateful to Víctor Cervantes for numerical experimentation
with the measures described in this paper, as well as for critically
reading the manuscript, and to Bárbara Amaral for sharing with us
her work. We thank the participants of the Purdue Winer Memorial Lectures
2018, especially Adán Cabello, for many fruitful discussions. We also
thank two anonymous reviewers for their insightful and constructive
comments.

\end{document}